\documentclass[aps,pra,twocolumn,nofootinbib]{revtex4-2}
\usepackage{amsmath,amssymb,amsfonts}
\usepackage{graphicx}
\usepackage{hyperref}
\usepackage{physics}
\usepackage{braket}
\usepackage{bm}

\begin{document}

\title{A Symmetry-Based Taxonomy of Quantum Algorithms}

\author{Sakshi Kumar}
\affiliation{School of AI, Amrita Vishwavidyapeetham, Delhi NCR}
\author{Sumit Chilkoti}
\affiliation{Department of Physics, Amrita Vishwavidyapeetham, Amritapuri}
\author{Mrittunjoy Guha Majumdar}
\email{Corresponding Author: mrittunjoy@dl.amrita.edu}
\altaffiliation{Also at the School of Natural Sciences and Engineering, National Institute of Advanced Studies}
\affiliation{School of AI, Amrita Vishwavidyapeetham, Delhi NCR}
\affiliation{Department of Physics, Amrita Vishwavidyapeetham, Amritapuri}

\begin{abstract}
We propose a taxonomy for quantum algorithms grounded in the fundamental symmetries—both continuous and discrete—underlying quantum state spaces, oracles, and circuit dynamics. By organizing algorithms according to their symmetry groups and invariants, we define distinct algorithm classes whose behavior, verification, and complexity can be characterized by the symmetries they preserve or exploit. This symmetry-centric classification not only reflects the deep connection between symmetries and conservation laws in physics, but also yields practical benefits for scalable and reliable quantum computation.
\end{abstract}

\maketitle

\textit{Introduction}. Symmetry plays a foundational role in the physical sciences, encoding conservation laws via Noether’s theorem, dictating selection rules, and underpinning the emergence of effective degrees of freedom \cite{gross1996,weinberg1995}. In quantum physics, symmetry governs dynamics, constrains Hamiltonians, and determines the structure of quantum states \cite{zee2016,cohen2019}. In the realm of quantum computing, similar principles apply: symmetries inform algorithm design \cite{childs2010,berry2015}, reduce circuit depth \cite{bravyi2022}, enable verification protocols \cite{xiong2021circuit}, and support error mitigation strategies through conserved quantities or symmetry-preserving subspaces \cite{bonet2018low,cai2021multi}. Despite these deep connections, a comprehensive classification of quantum algorithms grounded in the symmetries they exploit or preserve has not been developed. Previous taxonomies of quantum algorithms have been predominantly functional—organized by computational goal (for instance, simulation, search, optimization) or query structure (for instance, black-box models, hidden subgroup problems) \cite{montanaro2016,wang2022}. While such classifications provide utility, they fail to highlight the structural properties that make quantum algorithms both analyzable and verifiable. In contrast, symmetry-centered frameworks—commonplace in many-body physics and field theory—offer unifying abstractions that can guide both theoretical development and experimental implementation.
\\
\\
Recent advances in circuit symmetry verification \cite{xiong2021circuit}, symmetry expansion methods \cite{cai2021multi}, and zero-cost symmetry-based postprocessing \cite{bonet2018low} indicate that embedding and detecting symmetry in quantum circuits is not only feasible but advantageous. Moreover, symmetry constraints have been shown to reduce simulation complexity \cite{gottesman1997stabilizer}, facilitate error correction \cite{knill2005quantum,terhal2015quantum}, and optimize ansatz design in variational algorithms \cite{grimsley2019adaptive} and quantum machine learning models \cite{schuld2021machine,larocca2022group}. In this paper, we propose a symmetry-based taxonomy of quantum algorithms, wherein each algorithm class is defined by a fundamental symmetry group and its associated invariants. These include discrete groups such as $\mathbb{Z}_2$ and $S_n$, continuous Lie groups like $SU(2)$ and $SU(n)$, and more general algebraic symmetries such as tensor and oracle-invariant structures. Each class is examined through the lens of its group-theoretic structure, the induced decomposition of Hilbert space, and the practical consequences for algorithm efficiency, verification, and error resilience. This framework provides not only a conceptual reorganization of known quantum algorithms, but also a pathway for the principled design of new ones.
\\
\\
\textit{Algorithm Classes Defined by Symmetry}. Quantum algorithms can be naturally grouped into classes defined by the symmetry groups they preserve. Instead of summarizing these in tabular form, we now explore each symmetry-defined class in depth, highlighting the mathematical structure, representative algorithms, operational implications, and verification protocols. These classes provide not only structural constraints but also opportunities for more efficient algorithm design, simulation, and verification. Permutation-invariant algorithms are characterized by symmetry under the action of the symmetric group $S_n$, which permutes the indices of tensor product states in the Hilbert space. This arises naturally in quantum systems involving indistinguishable bosons or symmetrized inputs, such as in boson sampling with linear optics, where the indistinguishability of photon modes leads to computational complexity rooted in $S_n$-symmetry. Mathematically, $S_n$ acts by permuting the tensor components of an $n$-qubit state, and the relevant invariant subspaces are spanned by fully symmetrized basis vectors. These constraints reduce the effective Hilbert space dimension from $d^n$ to $\binom{n+d-1}{n}$, significantly simplifying both state preparation and measurement. Verification in this setting can be implemented through measurements of symmetrized observables and statistical tests of output permutation invariance. Cyclic-symmetric algorithms, governed by the group $\mathbb{Z}_n$ or its continuous analog $C_n$, exhibit invariance under modular addition or rotation. The group action takes the form $|x\rangle \mapsto |x + k \mod n\rangle$, and the structure of such algorithms is captured in the quantum Fourier basis, where each Fourier mode corresponds to a character of $\mathbb{Z}_n$. Canonical examples include Shor’s algorithm and other quantum Fourier transform (QFT)-based algorithms solving hidden subgroup problems. These symmetries enable the efficient implementation of modular arithmetic and are responsible for exponential speedups in structured search and periodicity detection. Verification protocols exploit the Fourier structure through phase estimation circuits and autocorrelation tests, allowing direct detection of symmetry-consistent behavior.
\\
\\
Rotationally symmetric algorithms are associated with continuous symmetries under special unitary groups such as $SU(2)$ and $SU(n)$. These arise naturally in the simulation of spin systems, Heisenberg models, and quantum walks on symmetric graphs. The mathematical formalism is grounded in the representation theory of Lie groups, where states decompose into irreducible representations labeled by angular momentum quantum numbers. Clebsch-Gordan decompositions further elucidate the structure of multi-qubit systems with rotational invariance. Operationally, such symmetries enable algorithms to restrict dynamics to conserved subspaces, such as those with fixed total spin projection $J_z$, thereby reducing resource overhead and increasing simulation fidelity. Verification can be performed by constructing projectors onto $SU(2)$-invariant sectors and checking the conservation of spin through commutation relations. Parity-conserving algorithms reflect invariance under $\mathbb{Z}_2$ transformations, corresponding to bit-flip or fermionic parity symmetries. This symmetry class is pervasive in quantum chemistry and fermionic simulation, where it plays a crucial role in simplifying Hamiltonians and reducing the number of variational parameters. In the Jordan-Wigner transformation, for example, parity structure allows the Hamiltonian to decompose into block-diagonal form. The mathematical structure relies on involutive generators such as Pauli operators or parity-check matrices, dividing the Hilbert space into even and odd subspaces. Qubit tapering and symmetry-based term elimination in variational quantum eigensolver (VQE) algorithms illustrate operational benefits. Verification is typically achieved by post-selecting on parity sectors and measuring stabilizer-like operators to ensure that parity is conserved during evolution.
\\
\\
Clifford-class algorithms are those restricted to the Clifford group, the normalizer of the Pauli group in the unitary group, which includes the Hadamard, phase, and CNOT gates. These algorithms play a central role in quantum error correction, state preparation, and classically simulable quantum computation. Under the stabilizer formalism, Clifford circuits preserve Pauli group structure, enabling efficient simulation via the Gottesman-Knill theorem. Examples include surface code syndrome extraction, teleportation protocols, and magic state distillation routines. Their operational utility lies in their error-robustness and low verification cost. Stabilizer measurements and Pauli frame tracking provide scalable and low-overhead verification techniques, allowing them to serve as trusted subroutines in fault-tolerant architectures. Oracle-symmetric algorithms represent a broad and powerful class in which symmetry is imposed not on the circuit structure, but on the invariance of a black-box function or oracle under some group action. These problems form the core of the hidden subgroup problem (HSP) family. Mathematically, a function $f: G \rightarrow S$ is said to be invariant under the left (or right) action of a subgroup $H \leq G$ if $f(g) = f(hg)$ for all $h \in H$. This leads to efficient quantum algorithms exploiting the group structure of the oracle, such as Simon’s problem with $\mathbb{Z}_2^n$ symmetry, or period finding in Shor’s algorithm with cyclic symmetry. Algorithmic speedups emerge from structured oracle access, and verification may be achieved through group-theoretic consistency checks and Fourier sampling to confirm that measured outcomes align with the underlying symmetry.
\\
\\
Finally, tensor-symmetric algorithms capture symmetries arising from automorphisms of tensor network or graph structures, relevant in quantum machine learning and quantum many-body simulations. These algorithms exploit geometric or combinatorial symmetries, where group actions commute with the network contraction structure and preserve entanglement patterns. Examples include equivariant quantum circuits used in geometric learning tasks, as well as symmetric tensor networks like MERA and PEPS in condensed matter simulations. Such symmetries drastically reduce the expressive capacity of variational ansätze to physically relevant subspaces and improve generalization in learning tasks. Verification techniques involve permutation-invariance tests on input-output data and correlation functions, ensuring consistency with the target symmetry. This symmetry-based classification of quantum algorithms offers a principled and unifying framework that reflects the underlying physics, guides efficient implementation, and supports robust verification. By grounding algorithmic behavior in group-theoretic structure, this taxonomy enables both practical and conceptual advances across quantum computation.
\\
\\
\textit{Symmetry Verification and Detection}. Verification of quantum algorithms is a critical challenge, particularly in the noisy intermediate-scale quantum (NISQ) era. The symmetry-based taxonomy proposed in this paper offers a powerful framework for structured verification, wherein each symmetry class implies distinct conservation laws, commutation relations, or subspace constraints that can be exploited to certify algorithmic correctness and detect errors. Below, we develop verification protocols tailored to each symmetry class. For permutation-invariant algorithms associated with the symmetric group $S_n$, verification can be carried out through measurements of symmetric observables that commute with all permutation operators. Specifically, testing invariance of output statistics under basis permutation provides a non-invasive check of the symmetrization constraint. In bosonic systems, detection of particle distinguishability errors can be achieved through statistical analysis of marginal distributions. Additionally, one may use projection operators onto the symmetric subspace, constructed from Young symmetrizers, to validate that the system remains within the intended $S_n$-invariant sector throughout computation.
\\
\\
Cyclic-symmetric algorithms based on $\mathbb{Z}_n$ or $C_n$ symmetries admit verification via Fourier-domain diagnostics. Because these algorithms often utilize the quantum Fourier transform (QFT) to access symmetry-adapted basis states, errors that break cyclic invariance manifest as anomalous phase distributions in the Fourier spectrum. Phase estimation circuits can be used to probe eigenvalues of shift operators, while autocorrelation functions provide signatures of broken periodicity. For algorithms like Shor's or hidden subgroup solvers, the detection of correct period structure directly confirms adherence to cyclic symmetry constraints. Rotationally symmetric algorithms, typically preserving $SU(2)$ or $SU(n)$ symmetries, benefit from verification via conserved quantities such as total spin or angular momentum projection. One approach involves inserting intermediate measurements of operators like $J_z$ or $J^2$, which commute with the system Hamiltonian, to ensure evolution remains confined to an invariant subspace. Clebsch–Gordan coefficients enable the construction of projectors onto irreducible representation sectors, allowing finer-grained verification. These methods are particularly effective in quantum simulation tasks where physical symmetries correspond directly to conserved observables.

Parity-conserving algorithms, governed by $\mathbb{Z}_2$ symmetry, support efficient verification through parity-check measurements. These often correspond to fermionic parity operators or global Pauli-$Z$ strings, whose eigenvalues must remain fixed throughout evolution. Post-selection on parity eigenvalues is a common NISQ technique to suppress errors and identify symmetry-breaking events. In variational algorithms, parity conservation can be enforced at the ansatz level, with violations indicating implementation or sampling errors. Stabilizer-like constructions, such as anticommuting parity-check operators, offer scalable and circuit-compatible tools for real-time verification. Clifford-class algorithms, structured by the Clifford group, enable some of the most efficient verification methods available. Because Clifford circuits map Pauli operators to Pauli operators under conjugation, the stabilizer formalism can be used to track quantum state evolution classically. This allows for runtime validation of circuit behavior through Pauli frame tracking. In fault-tolerant settings, syndrome measurements from surface codes or concatenated codes function as native symmetry detectors, verifying the preservation of code space symmetries. Furthermore, these circuits serve as trusted building blocks for bootstrapping verification in non-Clifford extensions, such as magic state distillation. 
\\
\\
Oracle-symmetric algorithms, which exploit invariance of a function $f: G \to S$ under group action, admit verification via group-theoretic tests. For example, in Simon’s algorithm, checking that $f(x) = f(x \oplus s)$ for candidate $s$ values confirms oracle symmetry. Fourier sampling over the group $G$ provides empirical evidence that output distributions conform to expected subgroup structures. Moreover, consistency tests—such as verifying closure and invariance under the hypothesized subgroup action—can be used to validate oracle behavior without full function reconstruction, leveraging the underlying symmetry for exponential savings. Tensor-symmetric algorithms, arising from graph or tensor network automorphisms, permit verification through structural invariants of the network. Input-output consistency under graph automorphisms or input permutations reveals whether equivariance is preserved. For instance, one may test whether correlation functions or entanglement spectra are invariant under known symmetry actions. In learning applications, invariance of model predictions under symmetric transformations of inputs serves as a diagnostic for architectural correctness. These techniques generalize classical model invariance tests to quantum settings and are critical for validating geometric and data-driven quantum circuits.
\\
\\
In all cases, symmetry verification provides a powerful toolkit for error detection and mitigation. Unlike full tomography, these methods scale favorably with system size, and their integration into circuit execution—either through ancilla-based subspace projections, measurement post-processing, or online statistical testing—makes them well-suited to contemporary quantum hardware. The use of symmetry as a verification principle thus extends the utility of these groups beyond classification and into the domain of practical quantum reliability.
\\
\\
\textit{Advanced Applications and Outlook}. The symmetry-based taxonomy of quantum algorithms presented in this paper opens a range of new avenues for advanced applications, design strategies, and theoretical inquiry. In this section, we outline how each symmetry class contributes to ongoing developments in quantum machine learning (QML), quantum simulation, and algorithmic complexity, while identifying key open questions at the intersection of symmetry, computation, and architecture. Symmetry principles play a central role in the development of efficient and generalizable quantum machine learning models. Tensor-symmetric algorithms, in particular, provide the mathematical foundation for equivariant QML architectures, where circuit ansätze are constrained to respect group actions such as permutation, rotation, or graph automorphism. This ensures that the learned model remains invariant or equivariant under transformations of input data, leading to improved sample efficiency and better generalization. Equivariant unitary layers derived from representations of $S_n$, $\mathbb{Z}_n$, or $SU(n)$ allow the systematic construction of neural-like quantum circuits whose expressivity is restricted to symmetry-respecting functions.
\\
\\
Permutation and cyclic symmetries are increasingly used to design kernel-based QML models that are invariant under permutations of features or time steps, enabling direct application to tasks such as molecule classification, time-series analysis, and graph-based learning. In these contexts, group-convolution techniques analogous to classical group-equivariant neural networks are being extended into the quantum domain. Rotationally symmetric ($SU(2)$, $SU(n)$) and parity-conserving ($\mathbb{Z}_2$) algorithm classes are foundational in quantum simulation of condensed matter, quantum chemistry, and lattice gauge theories. Simulations of spin systems, fermionic models, and gauge field dynamics naturally inherit these symmetries from the underlying Hamiltonians. Embedding these symmetries into the algorithmic structure allows for reduced resource requirements via subspace targeting, symmetry-constrained ansätze, and qubit tapering techniques. These strategies not only improve fidelity on NISQ devices but also reflect the physical correctness of simulations by preserving conserved quantities such as total spin or particle number.
\\
\\
Recent developments also point to hybrid architectures where Clifford-class circuits are used to initialize symmetry-respecting states or extract error syndromes, before applying symmetry-preserving non-Clifford unitaries for dynamics. This suggests a modular design approach to quantum simulation, enabled by symmetry segmentation. Oracle-symmetric algorithms, such as Simon’s algorithm or the abelian hidden subgroup problem, reveal how group structure can endow quantum algorithms with exponential advantage. These examples suggest a deeper connection between symmetry and computational complexity. The structure of the hidden subgroup directly governs query complexity and algorithmic performance, hinting at a classification of algorithmic power by the symmetry of the underlying oracle. Open problems include extending such frameworks to non-abelian groups, developing completeness notions for symmetry classes, and identifying general conditions under which symmetry grants quantum speedups.
\\
\\
Clifford-class algorithms also occupy a key position in complexity theory. They represent the boundary between classically simulable and universal quantum algorithms. Recent work explores how restricted symmetries—such as those stabilizing specific Pauli subgroups—affect computational power when combined with magic state injection or T gates, motivating further exploration of symmetry as a complexity resource. Several open problems emerge from the symmetry-based perspective. One is the classification of \textit{symmetry-breaking} quantum algorithms—those whose effectiveness depends on explicitly breaking a symmetry present in the problem instance. Understanding the resource implications of symmetry violation, particularly in variational settings, could inform ansatz design and training protocols.
\\
\\
Another frontier is the exploration of hierarchical or nested group structures. For instance, algorithms may preserve a subgroup of a larger symmetry group or transition across symmetry classes during computation. Developing a multiscale symmetry taxonomy that reflects such hierarchical embeddings could yield insights into algorithm dynamics, compilation strategies, and circuit optimization. Lastly, there is significant interest in formalizing the relationship between symmetry and quantum complexity classes. Can symmetry constraints be mapped to complexity class separations? Does the presence of a specific symmetry restrict or expand algorithmic expressivity in a provable way? These questions lie at the intersection of group theory, computational complexity, and quantum information theory, and answering them may offer a new organizing principle for quantum algorithm design. In summary, symmetries not only provide structure and constraints but also serve as computational resources. As quantum hardware matures, leveraging these symmetries for circuit compilation, verification, machine learning, and complexity analysis may be key to scalable and interpretable quantum computation.
\\
\\
\textit{Conclusion}. We have proposed a taxonomy of quantum algorithms grounded in the fundamental symmetries of their underlying Hilbert spaces, oracles, and circuit dynamics. By organizing algorithmic behavior through the lens of symmetry groups such as $S_n$, $\mathbb{Z}_n$, $SU(n)$, and the Clifford group, we have identified structural principles that govern design, verification, and complexity. This framework not only unifies diverse algorithm families under a common theoretical language but also provides practical tools for optimizing resource use, enhancing generalization in quantum machine learning, and enforcing physical constraints in simulation tasks. Moreover, it reveals how symmetries function as computational resources, with implications for error mitigation, expressivity control, and complexity class separations. Future research into symmetry breaking, group hierarchy, and the interplay between symmetry and quantum advantage may further solidify this approach as a foundational paradigm for scalable quantum software.
\\
\\
DECLARATIONS 
\\
\\
\textit{Availability of data and material} - All data and material referenced in this manuscript are publicly available and properly cited. No new datasets were generated during this work.
\\
\\
\textit{Competing interests} - The author declares no known competing financial interests or personal relationships that could have appeared to influence the work reported in this paper.
\\
\\
\textit{Funding} - This research received no specific grant from any funding agency in the public, commercial, or not-for-profit sectors.
\\
\\
\textit{Authors’ contributions} - 
MGM conceptualized the research framework and led the writing of the manuscript. SK and SC investigated various symmetries and their applications, and contributed to the writing and refinement of the manuscript. All authors reviewed and approved the final version of the manuscript.
\\
\\
\textit{Acknowledgements} - The authors did not receive assistance from any individual or organization in the preparation of this manuscript.
\\
\\
DATA AVAILABILITY
\\
\\
No data was generated or analyzed in this study.
\bibliographystyle{unsrt}
\bibliography{apssamp}

\end{document}